\documentclass[prl,twocolumn,showpacs,byrevtex,amsmath,amssymb,floatfix]{revtex4}
\usepackage{graphicx}
\usepackage{dcolumn}
\usepackage{bm}
\usepackage{amsmath}
\usepackage{gensymb}
\usepackage{mathtools}
\usepackage{xcolor}
\usepackage{ulem}

\begin{document}
\preprint{PREPRINT}

\title[Short Title]{Influence of Confinement on Flow and Lubrication Properties of a Salt Model Ionic Liquid Investigated with Molecular Dynamics}

\author{Miljan Da\v{s}i\'{c}}
\affiliation{
Scientific Computing Laboratory, Center for the Study of Complex Systems,
Institute of Physics Belgrade, University of Belgrade,
Pregrevica 118, 11080 Belgrade, Serbia
}

\author{Konstantinos Gkagkas}
\affiliation{
Advanced Technology Division, Toyota Motor Europe NV/SA,
Technical Center,
Hoge Wei 33B, 1930 Zaventem, Belgium
}

\author{Igor Stankovi\'{c}}
\email{igor.stankovic@ipb.ac.rs}
\affiliation{
Scientific Computing Laboratory, Center for the Study of Complex Systems,
Institute of Physics Belgrade, University of Belgrade,
Pregrevica 118, 11080 Belgrade, Serbia
}

\date{\today}

\begin{abstract}

We present a molecular dynamics study of the effects of confinement on the lubrication and flow properties of ionic liquids.
We use a coarse grained salt model description of ionic liquid as a lubricant confined between finite solid plates
and subjected to two dynamic regimes: shear and cyclic loading.
The impact of confinement on the ion arrangement and mechanical response of the system has been studied in detail and compared to static and bulk properties.
The results have revealed that the wall slip has a profound influence on the force built--up as a response to mechanical deformation and that at the same time
in the dynamic regime interaction with the walls represents a principal driving force governing the behaviour of ionic liquid in the gap.
We also observe a transition from a dense liquid to an ordered and potentially solidified state of the ionic liquid taking place under variable normal loads and under shear.

\end{abstract}
%

\maketitle
%

\section{Introduction}
In this work, the lubricating ability and flow properties of salt model ionic liquids (ILs) containing salt--like spherical cations and anions are studied.
ILs are molten salts typically consisting of large--size organic cations and anions.
The thermochemical stability, negligible vapor pressure, viscosity, wetting performance and other physicochemical properties of ILs are important factors contributing
to the interest in their research for lubricant applications~\cite{zhou2009ionic,hayes2010interface}.
In addition, their properties can be modified by an applied voltage using confining charged surfaces in order to build--up an electric field across the nanoscale film.
The applied potential can affect the structure of IL layers and lead to externally controllable lubricating properties~\cite{fajardo2015friction,fajardo2015electrotunable,capozza2015squeezout}.
There is also a significant flexibility in tuning the physical and chemical properties of ILs which can affect lubrication such as viscosity, polarity and surface reactivity,
by varying their atomic composition as well as the cation--anion combination.
The thermal stability and negligible vapor pressure of ILs enable their usage at a high temperature.

Regarding the ability of ionic liquids to dynamically penetrate between surfaces, i.e. wetting, sometimes it is considered that a low contact angle of the lubricant indicates
the affinity between the liquid and the surface, since the liquid is more likely to stay in the area in which it was initially placed.
It is also expected that a lubricant is going to penetrate into small--gap components.
However, the effect of wettability of the ionic liquids is not understood well.
The wetting of plate surfaces such as mica is known to be partial by at least some ILs~\cite{wang2013impact,beattie2013molecularly}.
Lubrication necessarily involves intimate molecular features of the liquid--solid plate interface, related with those mechanisms determining the ionic liquid's wetting of the plate.
When ILs are used as lubricants, their ions are ordered into layers and adsorbed onto surfaces~\cite{smith2013quantized}. These adsorption layers can reduce friction and wear, particularly
in the case of boundary lubrication~\cite{smith2013quantized}.

An important observation is that ILs confined between surfaces feature alternating positive and negative ionic layers, with an interlayer separation corresponding to the ion pair size~\cite{lubricants2013, tribint2017}.
However, determining the structure of ILs during flow and the mechanism of nanoscopic friction with ILs as lubricants, poses a great scientific challenge, and so far a few studies in this direction
have been performed~\cite{fajardo2015friction}.
ILs involve long--range Coulombic interactions inducing long--range order on far greater scales than the IL itself~\cite{mendonca2013ILmetal,VoeltzelC5CP03134F,CanovaC4CP00005F}.
Recent studies of IL lubricants~\cite{fajardo2015friction,fajardo2015electrotunable,capozza2015squeezout,tribint2017} have shown that if the molecules interact via non--bonded potentials (Lennard--Jones and Coulombic potential),
this can capture all main physical attributes of the IL--lubricated nanotribological system.
Therefore, molecular--scale simulations can provide important insights which are necessary for understanding the differences in flow behaviour between bulk and confined liquid lubricants and the mechanisms behind,
such as boundary layers formation in case of shearing and/or applied normal load.

For this study, we utilize our previously developed coarse grain molecular dynamics (MD) simulation setup consisting of two solid plates, and an ionic liquid lubricant placed between them~\cite{tribint2017}.
The motivation for the chosen values of relevant model parameters (i.e. velocities, pressures, temperatures, time duration of simulations) comes from potential applications of ILs as lubricants in automotive industry.
Under typical operation of internal combustion engines, the conditions inside the combustion chamber vary significantly.
Temperature can range from $300$~K to the values higher than $2000$~K, while pressure ranges from atmospheric to the values higher than $10$~MPa~\cite{holmberg2012carenergy}.
The piston reciprocates with a sinusoidal velocity variation with speeds varying from zero to over $20$~m/s, with a typical speed being around $1$m/s.
The time required for one revolution of the engine is of the order of $10^{-2}$~s, while the total distance travelled by the piston over this period is of the order of $0.2$~m.
Such scales are typically modelled using continuum mechanics simulations. However, such simulations cannot provide the physical insight which is necessary for understanding
the molecule--dependent processes that affect the tribological phenomena.
Therefore, we have impemented a coarse grain MD simulation setup which can, inter alia, provide useful insights to lubrication mechanisms of piston ring--cylinder liner contact in automotive engines.

The determination and design of new applicable lubricants require understanding of both general and specific behaviour of liquids when exposed to nanoscale confinement, shearing and normal load.
In this study our focus is on determing general features of ILs as nanoscale lubricants.
Hence, we have chosen the model of a generic IL which is simple in order to provide a wide perspective of relevant mechanisms governing the IL lubrication principles.

This paper is organized as follows: Section 2 introduces the model and MD simulation setup of the solids and lubricants used, while the motivation for the choices made is provided.
In Section 3 the structure and viscosity characteristics of the bulk ionic liquid are presented. Section 4 is dealing with the static and dynamic behaviour of confined IL.
It also presents the results of confined IL's shear behaviour. Section 5 includes an overview of the principal observations and conclusions.

\section{Model}
The model used in this work is a coarse--grained model of IL which has already been exploited in previous studies~\cite{fajardo2015friction,fajardo2015electrotunable,capozza2015squeezout,tribint2017}
and it is known as SM model (salt--like model). It is a charged Lennard--Jones system consisting of cations and anions.
There are two types of interatomic interactions in our system and both of them are non--bonded: Lennard--Jones (LJ) potential and Coulombic electrostatic potential.
In the current work we are comparing bulk and confined IL properties.
Therefore, there are three different atom types taken into consideration: $(i)$ cations, $(ii)$ anions and $(iii)$ solid plate atoms.
Between all types of atoms we apply full LJ 12-6 potential, with the addition of Coulombic electrostatic potential for the interactions between ions.
In our system the cations and the anions are charged particles, while the solid plate atoms are electroneutral.
Accordingly, we have implemented a LJ 12-6 potential combined with Coulombic electrostatic potential:

\begin{equation}
V\left(r_\mathrm{\textit{ij}}\right) = 4 \epsilon_\mathrm{\textit{ij}}
\left[\left(\frac{\sigma_\mathrm{\textit{ij}}}{r_\mathrm{\textit{ij}}}\right)^{12} -
\left(\frac{\sigma_\mathrm{\textit{ij}}}{r_\mathrm{\textit{ij}}}\right)^6\right] + \frac{1}{4 \pi \epsilon_0 \epsilon_r} \frac{q_iq_j}{r_\mathrm{\textit{ij}}}
\end{equation}

Parameters $\left\{\epsilon_\mathrm{\textit{ij}}, \sigma_\mathrm{\textit{ij}}\right\}$ define the LJ potential between different types of particles:
$i,j = {{\rm A},{\rm C}, {\rm P}}$ which refer to anions, cations and solid plate atoms, respectively.
The diameter of cations and anions is set to $\sigma_{\rm CC} = 5$~{\AA} and $\sigma_{\rm AA} = 10$~{\AA}, respectively.
The mass of cations and anions is $m_{\rm C} = 130$~g/mol and $m_{\rm A} = 290$~g/mol, respectively.
The asymmetry of ion sizes is typical in many experimentally explored systems and the parameters have already been explored in literature, cf. Ref.~\cite{capozza2015squeezout,tribint2017}.
The atoms of the solid plates have a diameter of $\sigma_{\rm PP} = 3$~{\AA}. The mass of the solid plate atoms is $m_{\rm P} = 65$~g/mol.
The LJ potential has a short--range impact, since it vanishes rapidly as the distance increases $\propto r^{-6}$,
while the Coulombic potential has a long--range impact, $\propto 1/r$. To handle long--range interactions, we have used a multi--level summation method (MSM)~\cite{hardy2009multilevel},
since it scales well with the number of ions and allows the use of mixed periodic (in $x$ and $y$ directions) and non-periodic (in $z$ direction) boundary conditions, which are present
in our simulation setup with confined IL. On the other hand, in our simulation setup with bulk IL, periodic boundary conditions are applied in all three directions ($\left\{x, y, z\right\}$).
Ions are modelled as coarse grain particles, the charge of which is set equal to elementary: $q_{\rm C} = + \it{e}$ and $q_{\rm A} = - \it{e}$, i.e., $\it{e} = \text{1.6} \cdot \text{10}^{\text{-19}}$~C.
The dielectric constant is set to $\epsilon_r = 2$ to account for the dielectric screening,
as in Refs.~\cite{fajardo2015electrotunable, capozza2015squeezout, tribint2017}.

In this study, modelling the elasticity of metallic plates plays a secondary role (central role belongs to the lateral and normal forces created by the lubricant).
Therefore, we have selected a simplified model in which plate atoms interact strongly with each other if they belong to the same plate, i.e., $\epsilon_{\rm PP} = 120$~kCal/mol,
as opposed, to a very weak interaction between the different plates $\epsilon_{\rm top/bottom} = 0.03$~kCal/mol.
The parameter $\epsilon_{\rm PP}$ is so strong in order to ensure that the initial configuration of the solid bodies will basically remain unchanged (apart from high frequency oscillations).
Furthermore, even though typical engineering systems are often metallic, our initial coarse grained MD studies of liquid behaviour according to the applied conditions
justified the implementation of a simpler solid system which does not feature substrate polarization, cf. Ref.~\cite{tribint2017}.
Finally, it is possible that the actual surfaces might feature carbon coatings or depositions, in which case the surface polarization can be of secondary importance.

In Table~\ref{tab:tabSM} we are presenting the values of $\{\epsilon_{ij}, \sigma_{ij}\}$ parameters used in our model.
Arithmetic mixing rules for the LJ parameters are applied: $\epsilon_{ij} = \sqrt{\epsilon_i \cdot \epsilon_j}, \sigma_{ij} = \left(\sigma_i + \sigma_j\right)/2$.

\begin{table}
\begin{center}
\begin{tabular}{ |c|c|c| }
 \hline
 pair $ij$&  $\epsilon_{ij}$~[kCal/mol] & $\sigma_{ij}$~[{\AA}] \\
 \hline
 \hline
 CC & 0.03 & 5 \\
 \hline
 AA & 0.03 & 10 \\
 \hline
 CA & 0.03 & 7.5 \\
 \hline
 PC & 0.3 & 4 \\
 \hline
 PA & 0.3 & 6.5 \\
  \hline
 PP & 120 & 3 \\
 \hline
\end{tabular}
\end{center}
\caption{List of LJ parameters used in simulations.}
\label{tab:tabSM}
\end{table}

\section{Bulk Ionic Liquid}
All MD simulations in this study were performed using the LAMMPS software~\cite{plimpton1995fast}.
The bulk ionic liquid is implemented by randomly placing a chosen number of ions ($N_{\rm C} = N_{\rm A} = 1000$) into a 3D simulation box that is periodic in all three directions.
In order to make the bulk IL comparable with its confined counterpart, we have chosen a simulation box volume which enables the pressure experienced by the confined IL.
More specifically, for the present system, the pressure is $p \approx 1$~MPa.
The Nose--Hoover NVT thermostat was used to control the temperature and was set to $T = 330$~K. The system was relaxed for $t_{\rm tot} = 3 \cdot 10^7$~fs until the internal energy had converged and the pressure had approached
the desired value. The simulation timestep was $dt = 0.5$~fs. We have obtained pressure stabilization at $\left\langle p \right\rangle = 1.1$~MPa with a side length of the cubic simulation box at $L = 99$~{\AA}.
The energy relaxed to a value of $\left\langle E_{\rm int} \right\rangle = 0.7597$~kCal/mol.
The molar and mass density of the bulk IL is $\rho_{\rm n} = 3400$~$\text{mol}/\text{m}^{\text{3}}$ and $\rho_{\rm m} = 719$~$\text{kg}/\text{m}^{\text{3}}$ respectively.

\begin{figure}[htb]
\includegraphics[width = 8 cm]{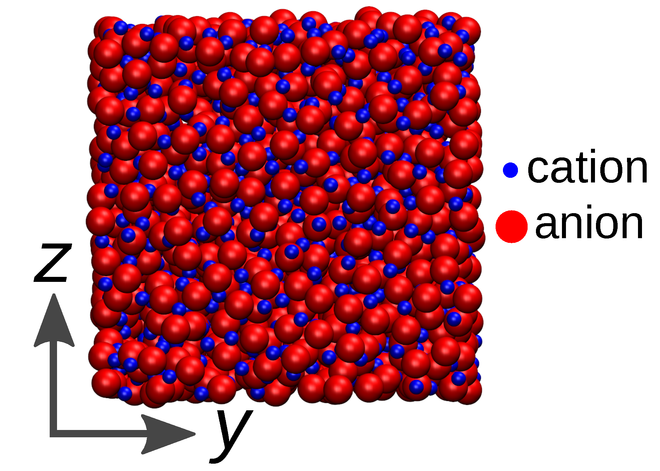}
\caption{Configuration snapshot ($yz$ cross--section) of a bulk IL at the end of relaxation simulation.
Cations are represented as smaller blue spheres and anions as larger red spheres.
}
\label{fig:bulkIL_SM}
\end{figure}

\begin{figure}[htb]
\includegraphics[width = 8 cm]{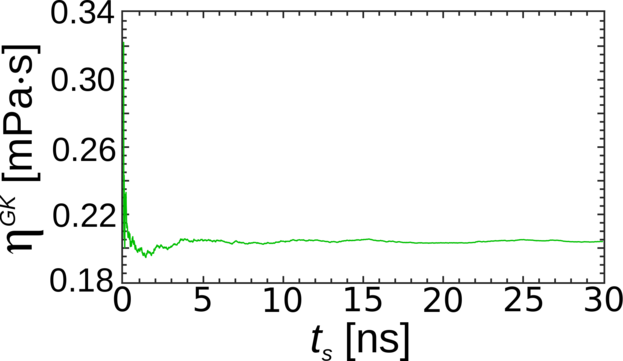}
\caption{Dependence of Green--Kubo (GK) viscosity coefficient $\eta^{\rm GK}$ on simulation time $t_s$ in case of bulk ionic liquid.
The time needed to obtain the viscosity coefficient is around $t_{\rm rel} = 5$~ns.
}
\label{fig:viscosity_timeSM}
\end{figure}

\begin{figure}[htb]
\includegraphics[width = 8 cm]{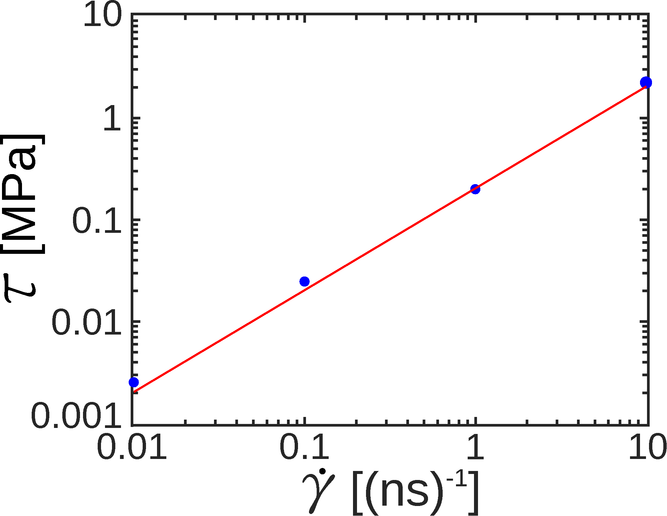}
\caption{Average stress tensor component $\tau$ in function of the shear rate $\dot{\gamma}$ of a bulk SM ionic liquid.
We have conducted shearing simulations on four orders of magnitude of the shear rate $\dot{\gamma}$, therefore with three orders of magnitude span,
which is followed by three orders of magnitude span of $\tau$. Points are obtained via shearing simulations and solid line is obtained according to:
$\tau = \eta^{\rm GK} \cdot \dot{\gamma}$, where $\eta^{\rm GK}$ is obtained via Green--Kubo relation.
}
\label{fig:tau_shear_etaGK}
\end{figure}

\begin{figure*}[htb]
\includegraphics[width = 16 cm]{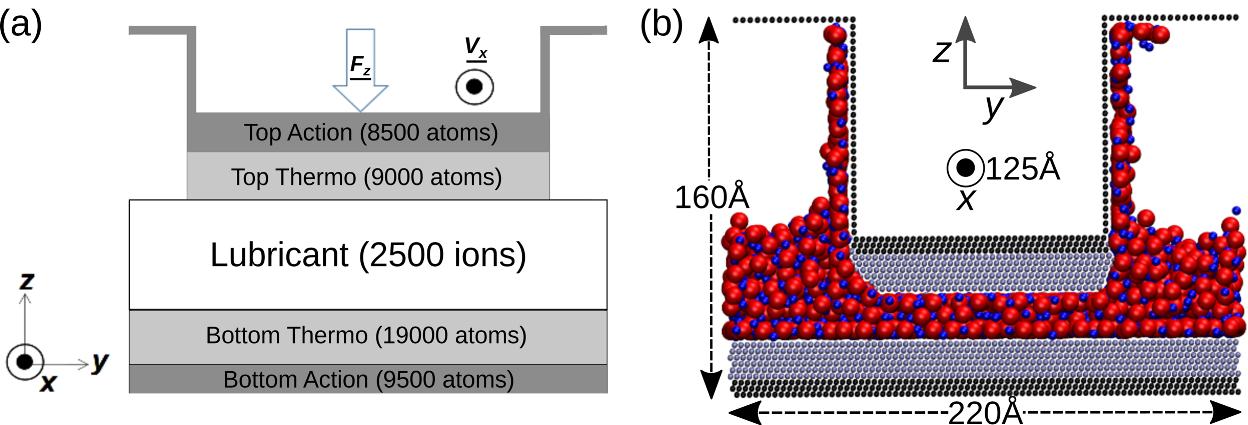}
\caption{(a) Schematic of the simulation setup shown as $yz$ cross--section.
There are two solid plates at the top and bottom of the system (i.e. Top and Bottom plate, names are chosen according to their position along the $z$ axis), consisting of two regions:
at the outermost ones the particles are moving as a single entity (Top/Bottom Action)
and at the innermost ones the particles are thermalized at a controlled temperature (Top/Bottom Thermo).
The thermalized layers are in direct contact with the lubricant while the action layers are used
to impose external velocity and/or force to the solid plates. (b) Front ($yz$) view with respect to
the shear direction. A typical simulation configuration and key dimensions of the geometry are given.
The solid plates are made up of FCC (111) atomic layers. The ionic liquid is composed of an equal
number of cations and anions (cations: smaller blue spheres; anions: larger red spheres).
}
\label{fig:schemeSM}
\end{figure*}

We have calculated the viscosity in two ways: using the Green--Kubo relation since the viscosity of a system can be represented as an integral of the autocorrelation function~\cite{viscardy2007transport},
and using non--equilibrium molecular dynamics simulations with different shear strains.

In the non--equlibrium shearing simulations, the bulk IL is placed into a triclinic (non--orthogonal) simulation box with periodic boundary conditions applied in all three directions.
Due to the deformation of the simulation box, every point in the box has an additional velocity component (a so called \textit{streaming velocity}).
In order to prevent the streaming velocity from affecting the thermal kinetic energy, we use the so-called \textit{SLLOD} thermostat~\cite{evans1984nonlinear,daivis2006simple}.
The \textit{SLLOD} thermostat accounts for the streaming velocity which depends on an atom's position within the simulation box and it needs to be accounted for controlling the temperature.

Controlled shearing of the simulation box results in a stress acting on IL, which is quantified via the stress tensor.
The relation between the stress tensor $\tau_{ij}$ components and the shear rate $\dot{\gamma}_{ij}$ of corresponding shear strain $\epsilon_{ij}$, with coefficient of viscosity $\eta_{ij}$ as a proportionality constant is:
$\tau_{ij} = \eta_{ij} \cdot \dot{\gamma_{ij}}$ where $ij = \left\{xy, xz, yz\right\}$. We have applied three independent shear strains ($\epsilon_{xy}, \epsilon_{xz}, \epsilon_{yz}$).
For each of them we have calculated its corresponding stress tensor component ($\tau_{xy}, \tau_{xz}, \tau_{yz}$).
All shear strains were the same: $\epsilon_{xy} = \epsilon_{xz} = \epsilon_{yz} = \epsilon = 1$ leading to the shear rate of $\dot{\gamma} = \epsilon \cdot \frac{1}{t_{\rm tot}} = \frac{1}{t_{\rm tot}}$,
where $t_{\rm tot}$ is the total simulation time of the shearing simulations. We have performed simulations at four orders of magnitude of the total simulation time: $t_{\rm tot} = \left\{0.1, 1, 10, 100\right\}$~ns,
and thus at four orders of magnitude of the corresponding shear rate.
In this way we wanted to check the quality of our relaxation procedure and if there are shear rate dependence changes in the system.
We have iterated the shearing simulations (at a shearing velocity of $1$~m/s) using the output of the previous run as the input of the next run, obtaining higher strains (up to a strain of $5$).
We did not observe a strain dependence in the response of the system, meaning that the result is unaffected if the strain is further increased.

In Figure~\ref{fig:viscosity_timeSM}, we show the time relaxation of the Green--Kubo viscosity coefficient, which stabilizes around $\eta^{\rm GK} = 0.2039$~$\text{mPa} \cdot \text{s}$.
The configuration snapshot of the bulk IL at the end of the simulation (cf. Figure~\ref{fig:bulkIL_SM}) shows that the ions remain randomly positioned, like they were at the start of simulation,
which indicates the liquid state of the bulk ionic liquid. The simulations for all three shear strains give similar values of stress components, and resulting values are shown in Figure~\ref{fig:tau_shear_etaGK}.
The points $\left\{\dot{\gamma}, \tau\right\}$ were obtained via shearing simulations and the solid line was obtained according to $\tau = \eta^{\rm GK} \cdot \dot{\gamma}$, where $\eta^{\rm GK}$ was obtained
via Green--Kubo relation.
Hence, we conclude that the results of shearing simulations are in agreement with the results of relaxation simulation and therefore there are no changes in the bulk system which are shear rate dependent.

\section{Confined Ionic Liquid}
In order to study the properties of our ionic liquid under confinement, we use a setup consisting of two solid plates (so called Top and Bottom plate)
and ionic liquid lubricant placed between them. Such simulation setup has been introduced and described in detail in our previous paper~\cite{tribint2017}, hence at this point we will describe it briefly.
The geometry is shown as a schematic in Figure~\ref{fig:schemeSM}(a) together with the number of the coarse grain particles used.
In Figure~\ref{fig:schemeSM}(b) we show a configuration snapshot of our system in $yz$ cross--section.
By implementing such a geometry we have attempted to achieve a realistic particle squeeze--out behaviour with the formation of two lateral lubricant regions in a similar manner
to the simulations of Capozza et al.~\cite{capozza2015squeezout}. For the description of the solid surfaces we have combined rigid layers of particles moving as a single entity
on which the external force or motion is imposed, denoted by "Top Action" and "Bottom Action" in Figure~\ref{fig:schemeSM}(a), with thermalized layers, denoted by "Top Thermo" and "Bottom Thermo"
in which particles vibrate thermally at $T = 330$~K. The particles in the Top and Bottom action layers are moved as rigid bodies and particles in the thermo layers are allowed
to move thermally. In this way, we prevent a progressive deformation of the plates during the cyclic movement. The thermo layers only vibrate thermally since a strong LJ interaction holds them together.
The ionic liquid is neutral in total, so the total number of cations and anions is the same: $N_{\rm C} = N_{\rm A} = N_{\rm IL}/2$. In the present simulations the total number of IL atoms is $N_{\rm IL} = 2500$.

The plates are driven along the $x$ direction at a constant velocity $V_x$, as shown in Figure~\ref{fig:schemeSM}(a).
The solid plates are made up of nine atomic layers at a FCC $\left(111\right)$ lattice arrangement. Periodic boundary conditions are applied in the $x$ and $y$ directions,
while the simulation box is kept fixed in the $z$ direction.
The Bottom plate can therefore be considered to be infinite, while the Top plate is surrounded by the lateral reservoirs, in which the lubricant can freely expand.
The lateral reservoirs are implemented as a mechanistic way for allowing the lubricant to be dynamically squeezed in or out as an external load or velocity is applied.
The number of lubricant molecules effectively confined inside the gap can dynamically change depending on the loading conditions.
This is important for exploring the possible states of a mechanical system of particles that is being maintained in thermodynamic equilibrium (thermal and chemical)
with a lubricant reservoir (i.e., void spaces in tribological system).
The nano--tribological system is open in the sense that it can exchange energy and particles, realizing an effectively grand--canonical situation~\cite{GaoPRL1997,GaoJPCB2004}.

We have shown that our bulk IL is a Newtonian fluid: the validity of $\tau = \eta^{\rm GK} \cdot \dot{\gamma}$ relation over the whole range of shearing rate $\dot{\gamma}$ supports that fact.
Our model does not assume the nature of viscous response of IL. Only based on simulation results we conclude that bulk salt model (SM) IL behaves as a Newtonian fluid.
For a different choice of parameters one could obtain power law or solid like behaviour.
On the other hand, confinement has a profound influence on the structure of ILs in thin films~\cite{lubricants2013,GaoPRL1997,PerkinPCCP2012,tribint2017},
therefore when the same IL is confined it does not behave as a Newtonian fluid, as we will show in the rest of the paper.

\begin{figure}[htb]
\includegraphics[width = 8 cm]{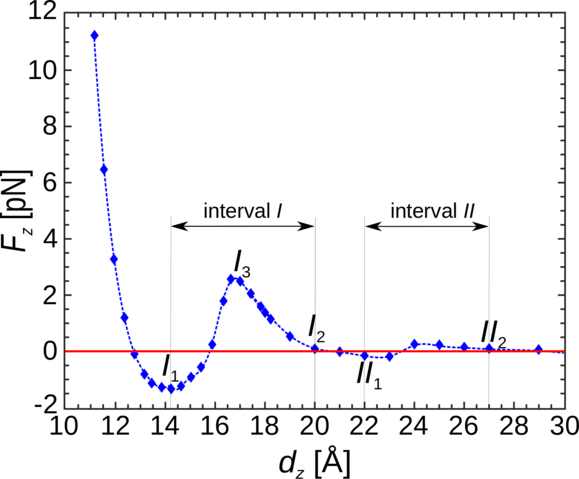}
\caption{Dependence of normal force $F_{z}$ acting on the Top plate on plate-to-plate distance $d_{z}$.
Five characteristic points \{\textit{I}$_1$, \textit{I}$_2$, \textit{I}$_3$, \textit{II}$_1$, \textit{II}$_2$\} with corresponding interplate distances $d_z \approx \left\{14, 20, 17, 22, 27\right\}$~{\AA} are marked
on the $F_z\left(d_z\right)$ curve. Also, the two characteristic intervals of $d_z$ are labeled, where the interval \textit{I} is bounded by the points \textit{I}$_1$ and \textit{I}$_2$,
while the interval \textit{II} is bounded by the points \textit{II}$_1$ and \textit{II}$_2$.
The horizontal solid line denotes $F_z = 0$~pN. The dashed line connects the points obtained from the simulation and serves as a visual guide.
}
\label{fig:fd_static}
\end{figure}

The confining surfaces can induce ordering of the particles in their vicinity. We have used simulations to obtain the static force--distance characteristic~\cite{tribint2017}.
In order to determine a reliable static force--distance characteristic, at each calculation point we have to ensure the system is in equilibrium.
Concerning the realization of those simulations the inter--plate gap is modified in the following manner:
the gap is increased or decreased (i.e., the Top--Bottom plate distance is changed) with a constant velocity $V_z = 5$~m/s for a \textit{move} period of time $t_{move} = 20$~ps;
thereafter, we apply conjugated gradient minimization on the ions in order to minimize their internal energy and relax them after the \textit{move} period.
As the energy minimization is performed, the ions take positions which ensure their minimal internal energy and the Top plate stays fixed for a \textit{stay} period of time $t_{stay} = 50$~ps,
during which period the average value of the normal force is calculated; that value is presented as a simulation point in $F_z\left(d_z\right)$ static characteristic, cf. Figure~\ref{fig:fd_static}.
In order to avoid systematic errors due to the initial position or direction, the plate movement is performed in different directions and from different initial configurations,
hence the Figure~\ref{fig:fd_static} shows the averaged values of several realizations.

\begin{figure*}[htb]
\includegraphics[width = 16 cm]{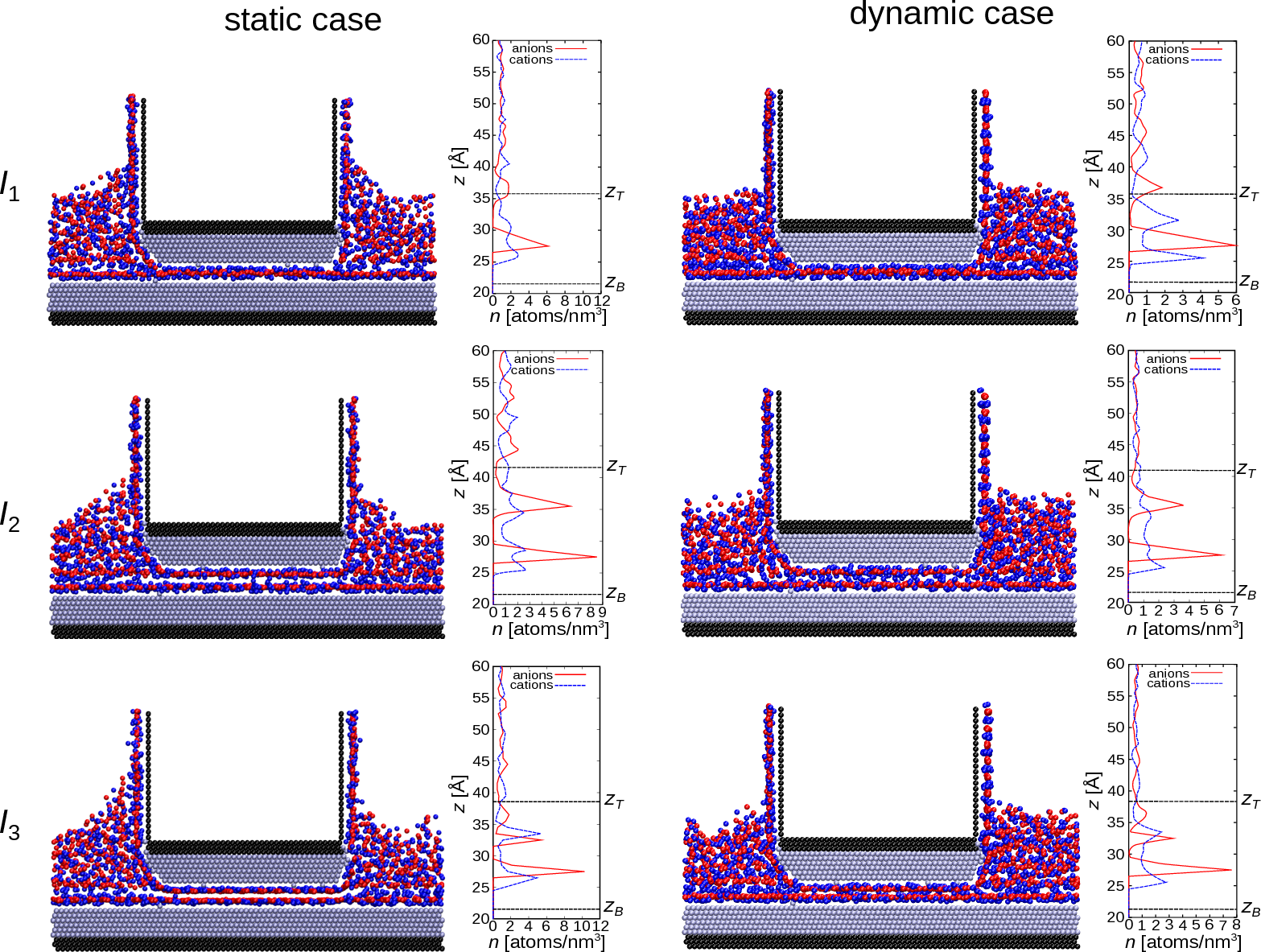}
\caption{Configuration snapshots ($yz$ cross--section) accompanied with ionic density distribution along the $z$ direction in three representative points of the interval \textit{I}:
\{\textit{I}$_1$, \textit{I}$_2$, \textit{I}$_3$\}. Left panels correspond to the static case of Top plate's movement, while right panels correspond to the dynamic case of Top plate's movement.
}
\label{fig:conf_interval1}
\end{figure*}

\begin{figure*}[htb]
\includegraphics[width = 16 cm]{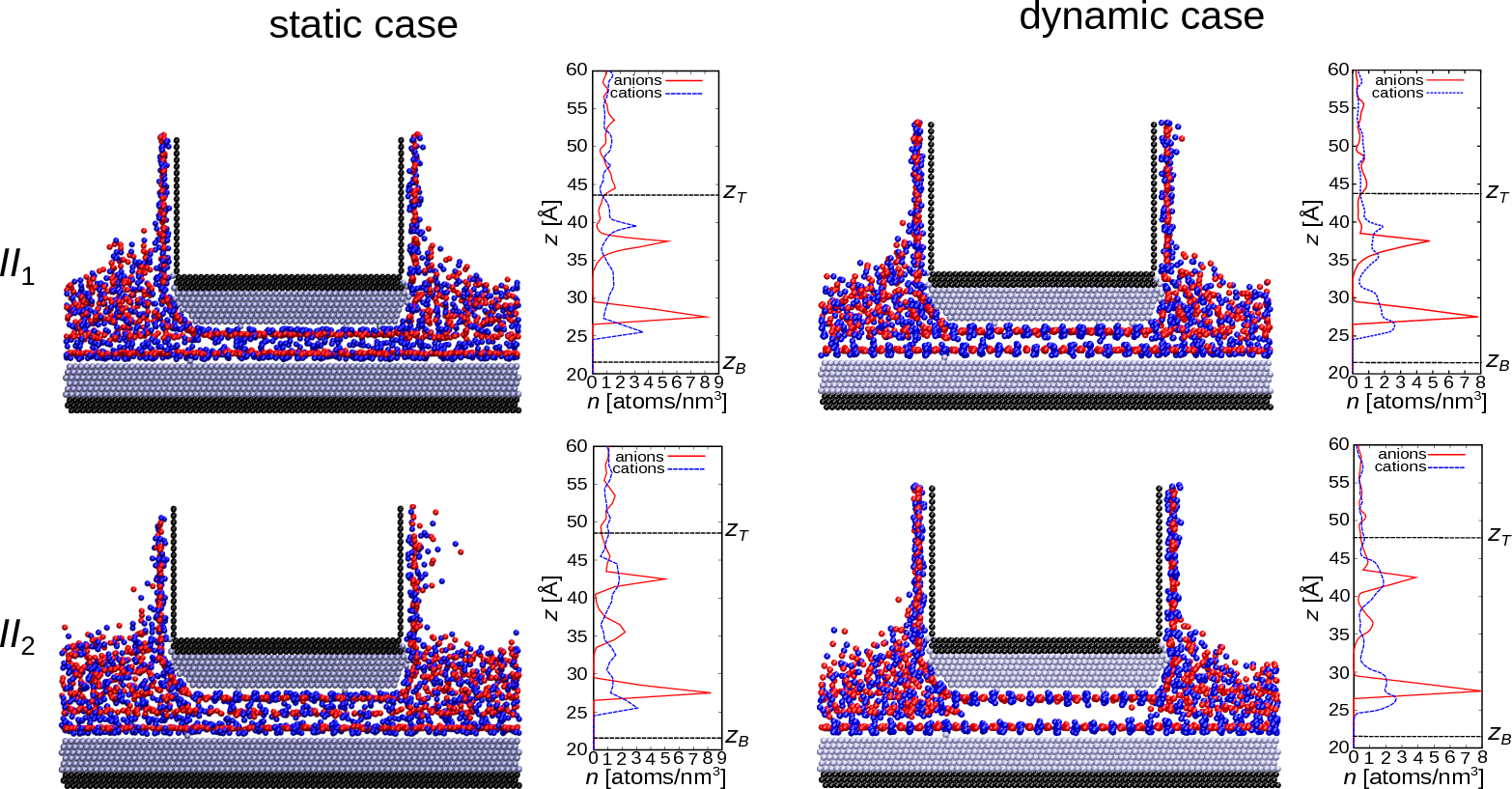}
\caption{Configuration snapshots ($yz$ cross--section) accompanied with ionic density distribution along the $z$ direction in two representative points of the interval \textit{II}: \{\textit{II}$_1$, \textit{II}$_2$\}.
Left panels correspond to the static case of Top plate's movement, while right panels correspond to the dynamic case of Top plate's movement.
}
\label{fig:conf_interval2}
\end{figure*}

\begin{figure}[htb]
\includegraphics[width = 8 cm]{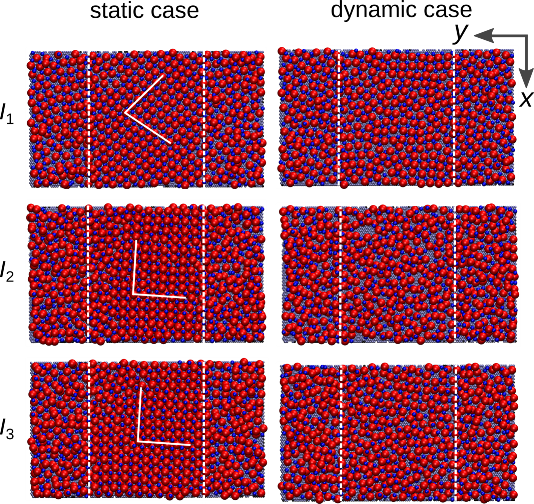}
\caption{Configuration snapshots ($xy$ cross--section) in three representative points of the interval \textit{I}: \{\textit{I}$_1$, \textit{I}$_2$, \textit{I}$_3$\}.
Left panels correspond to the static case of Top plate's movement, while right panels correspond to the dynamic case of Top plate's movement.
We have highlighted the confined region with dashed lines (Top plate's width along the $y$ axis is a half of the total system's width) and also we have sketched crystallization patterns with solid lines.
Periodic boundary conditions are applied in the $x$ and $y$ directions, while simulation box, which is cubic, is kept fixed in the $z$ direction.
}
\label{fig:confxyview_interval1}
\end{figure}

\begin{figure}[htb]
\includegraphics[width = 8 cm]{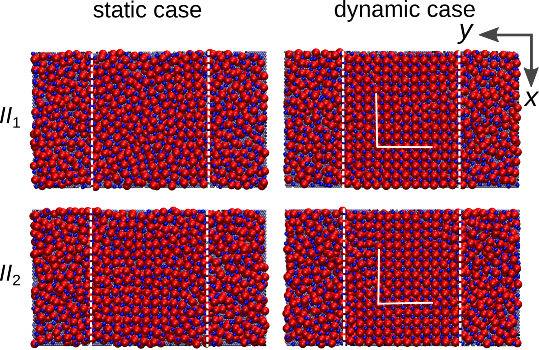}
\caption{Configuration snapshots ($xy$ cross--section) in two representative points of the interval \textit{II}: \{ \textit{II}$_1$, \textit{II}$_2$\}.
Left panels correspond to the static case of Top plate's movement, while right panels correspond to the dynamic case of Top plate's movement.
We have highlighted the confined region with dashed lines (Top plate's width along the $y$ axis is a half of the total system's width) and also we have sketched crystallization patterns with solid lines.
}
\label{fig:confxyview_interval2}
\end{figure}

\subsection{Equilibrium Behaviour of Confined Ionic Liquid}
A non--monotonous behaviour of the normal force $F_z$ acting on the Top plate can be observed in Figure~\ref{fig:fd_static} as the plate-to-plate distance is changing from one point at which system is equilibrated
to another, using the previously described procedure.
The points $\left(d_z, F_z\right)$ have been obtained through our simulations, while the dashed line serves as a visual guide.
It can be seen that the normal force strongly depends on the inter--plate distance.
The presence of negative values of normal force $F_z$ can be understood as the IL trying to reduce the plate-to-plate distance due to adhesion phenomena. These changes of the normal force are correlated
with the extraction and inclusion of IL layers into the gap, as already observed experimentally, cf. Ref.~\cite{lubricants2013}. During the performed stationary state simulations,
the cationic--anionic layers were squeezed out in pairs, in order to keep the system locally neutral, as observed in experimental studies~\cite{lubricants2013, GaoPRL1997, PerkinPCCP2012, hayes2011double, smith2013quantized}.

There is a strongly decreasing trend of the maximal normal force which can be sustained by the system as the number of ionic layers confined between the plates increases, i.e., for the two ionic layers
the maximal force $F^{\rm I}_{z,max} = 3$~pN, while for the three ionic layers it is $F^{\rm II}_{z,max} = 0.25$~pN. In our model, the Lennard-Jones interaction between the plates and the ions is ten times stronger
than between the ions themselves. The ionic layers closest to the plates are therefore more stable than the layers in the midpoint of the gap (interval {\it II}).
As a result, the three-layer system becomes less dense and can build up a lower normal force compared to the two-layer system. We have selected two intervals of interest for the interplate distance
which capture the presence of local maxima and subsequent minima of the normal force $F_z$ accompanied with the compression of IL. This corresponds to the expulsion of a cation--anion layer pair from the gap.
The intervals are: $d^{I}_z = \left[14.2, 20\right]$~{\AA}, $d^{II}_z = \left[22, 27\right]$~{\AA}, and they are labeled as \textit{I} and \textit{II} respectively.
In order to understand the changes of the system configurations and to correlate them with the changes of the inter--plate distance, snapshots of the system from the MD simulations corresponding to
several characteristic points of the intervals \textit{I} and \textit{II} have been selected and studied in more detail: \textit{I}$_{1,2}$, \textit{II}$_{1,2}$ which correspond to the limits of the intervals,
and the local maximum of the interval \textit{I}, labeled as $I_3$.

The left vertical panel of Figure~\ref{fig:conf_interval1} shows the system configuration in the $yz$ cross-section and the ionic density distribution along the z--direction obtained in the equilibrium force--distance
simulations for the three characteristic points of the interval \textit{I}: \{\textit{I}$_1$, \textit{I}$_2$, \textit{I}$_3$\}, corresponding to plate-to-plate distances $d_z = \left\{14.2, 20, 17.2\right\}${\AA}, respectively.
In Figure~\ref{fig:conf_interval2} the left vertical panels show analogous results for the two characteristic points of the interval \textit{II}: \{\textit{II}$_1$, \textit{II}$_2$\}, corresponding to
plate-to-plate distances $d_z = \left\{22, 27\right\}${\AA}, respectively.
The ions are depicted smaller than their LJ radii in order to allow a direct observation of the layering. The positions of the atomic centres of the innermost atomic layers of the Top and Bottom plate
are labeled in Figures~\ref{fig:conf_interval1} and~\ref{fig:conf_interval2} as $z_{\rm T}$ and $z_{\rm B}$ respectively. As the Bottom plate is kept fixed during the whole simulation,
$z_{\rm B}$ remains constant while $z_{\rm T}$ changes with the Top plate displacement. A general feature observed under all conditions was the formation of cationic layer close to the plates.
The reason is the smaller size of the cations ($\sigma_{\rm CC} = 5${\AA}) compared to the anions ($\sigma_{\rm AA} = 10${\AA}). Following, the second layer gets induced by the first one (due to Coulombic interaction)
and it is populated by anions. The distance between the first and the second layer from the bottom is in the range of $1-2.5$~{\AA}, meaning that while the centres of mass of the particles are in different layers,
the layers themselves overlap as their distance is smaller than the particle diameters.

From Figure~\ref{fig:conf_interval2} we observe that the anionic monolayer thickness is roughly $7${\AA}
and corresponds to $10/\sqrt{2}${\AA}, i.e., the anions are placed in the centers of the squares formed by the cations of the neighboring layers (the diameter of an anion is $10${\AA}).
In addition to the $yz$ cross--section configuration snapshots together with the ionic density distribution along the $z$ axis, shown in the left panels of Figures~\ref{fig:conf_interval1}
and~\ref{fig:conf_interval2} for the cases of intervals \textit{I} and \textit{II}, respectively, we have prepared the $xy$ cross--section configuration snapshots,
shown in the left panels of Figures~\ref{fig:confxyview_interval1} and~\ref{fig:confxyview_interval2}.

\begin{figure}[htb]
\includegraphics[width = 8 cm]{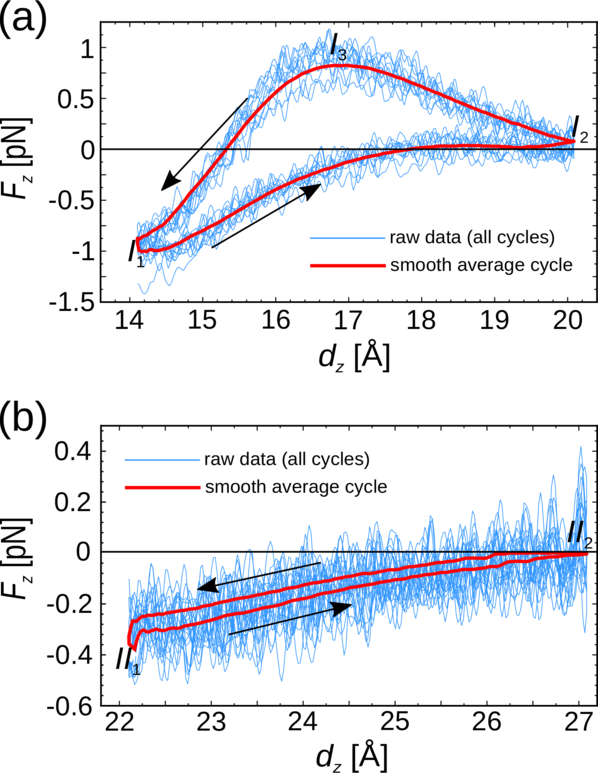}
\caption{This figure presents the results of dynamic extension--compression cycles in the intervals \textit{I} and \textit{II}.
In panel (a) we present dynamic $F_z\left(d_z\right)$ characteristic in the interval \textit{I}: thin lines represent the hystereses of ten dynamic cycles, solid line on top of them is the smooth average hysteresis.
There is also a solid horizontal line which corresponds to $F_z = 0$.
Panel (b) is analogous to the panel (a), just it presents the results in the interval \textit{II}.
}
\label{fig:fd_dynamic}
\end{figure}

\subsection{Cyclic Compression of Confined Ionic Liquid}
We have investigated the dynamic behaviour of the IL during a periodic linear movement of the Top plate along the $z$ axis, between the two limiting points of the intervals \textit{I} and \textit{II}.
The space between the solid plates was in this way periodically expanded and compressed.
Periodic movements of the Top plate were performed at three constant velocities $V_z = \left\{0.1, 1, 10\right\}$~m/s but no velocity dependent differences in the system behaviour were observed.
We have performed ten cycles in order to determine how much the cycles differ and to determine a statistically reliable average cycle.
The confined ionic liquid lubricant responds to the cyclic movement with a hysteresis in the normal force $F_z\left(d_z\right)$ as shown in Figure~\ref{fig:fd_dynamic}.
We present both raw data of all cycles (thin solid lines) and a smooth average cycle (thick solid line).
In the case of interval \textit{I} there are three points of interest \{\textit{I}$_1$,\textit{I}$_2$,\textit{I}$_3$\}, corresponding to the points noted in Figure~\ref{fig:fd_static}.
Points \textit{I}$_1$ and \textit{I}$_2$ are the starting and ending point respectively and the point \textit{I}$_3$ corresponds to the maximum of the normal force $F_z$ in the smooth average cycle.
We observe that between each two of those points there are clear tendencies in the average cycle of the normal force $F_z\left(d_z\right)$.
First, in the segment \textit{I}$_1\rightarrow$\textit{I}$_2$, i.e. in the extension half of the cycle, there is a continuous increase of normal force $F_z$ from negative values up to the value around zero in point \textit{I}$_2$.
In point \textit{I}$_1$ there is one anionic layer confined in the gap and normal force $F_z$ has a negative value.
With the dynamic increase of the gap ions are pulled--in from lateral reservoirs into the gap.
In point \textit{I}$_2$ an additional cationic--anionic layer pair is fully formed in the gap, hence increasing the number of confined anionic layers to two.
Next, there is the segment \textit{I}$_2\rightarrow$\textit{I}$_3$ where the ions are compressed within the gap, which is consistent with the continuous increase of normal force $F_z$.
In this segment, the normal force $F_z$ takes positive values meaning that the ionic liquid shows resistance to the compression but does not flow out.
After that, in segment \textit{I}$_3\rightarrow$\textit{I}$_1$ there is a sharp decrease of normal force $F_z$ which is correlated with the squeezing--out of the additional cationic-anionic layer taken in
from the lateral reservoirs during the extension half--cycle. During the compression half--cycle there is a return to the initial state \textit{I}$_1$, where the gap contains one compact anionic layer.

We should note that the distributions of cations and anions in the dynamic case for interval \textit{I} bear close resemblance.
Let us now discuss the changes in the number of confined ionic layers as a function of the inter--plate distance and correlate them with the changes in the normal force $F_z$ acting on the Top plate:
in the range $d_z = \left[11, 14.2\right]$~{\AA} the normal force $F_z$ acting on the Top plate has a steep decrease, reaching the minimum at point \textit{I}$_1$.
For the point \textit{I}$_1$ at $d_z = 14.2$~{\AA}, cf. Figure~\ref{fig:conf_interval1}, we can observe a pronounced peak in the anion density distribution which is aligned with a well--defined anionic layer inside the gap.
In the case of cations, there are two peaks attached below and above the anionic peak. This situation corresponds to the formation of two incomplete cationic layers inside the gap.
The value of normal force $F_z$ is negative and in point \textit{I}$_1$ it has the deepest minimum when considering the whole $F_z\left(d_z\right)$ characteristic.
With increasing plate-to-plate distance $d_z$ the normal force $F_z$ is increasing, with a sign change of normal force $F_z$ around $d_z = 15.7$~{\AA} in the equilibrium case and $d_z = 17.8$~{\AA} in the dynamic case,
cf. Figure~\ref{fig:fd_static} and ~\ref{fig:fd_dynamic}(a), respectively. This means that before this point the IL is pulling the plates together, since the ions strive to reduce their interlayer distance.
After this point, for $F_z > 0$, enough ions are pulled inside the gap and the IL now pushes the plates apart. Such behaviour is typically observed in systems exhibiting layering transition,
already seen in systems of both neutral molecules ~\cite{bhushan1995nanotribology} and ILs~\cite{lubricants2013}.
With reversing into compression in Figure~\ref{fig:fd_dynamic}(a), the normal force $F_z$ reaches a local maximum in the point \textit{I}$_3$ at $d_z = 17.2$~{\AA}.
This is actually the location of the maximum in the equilibrium case as well. With the further decrease of $d_z$ beyond the point \textit{I}$_3$ there is a continuous decrease of the normal force up to
the distance $d_z = 14.2$~{\AA} as IL starts flowing out of the gap. Still, one should note that there are two differences between the two systems:
($i$) the sign of the normal force in point \textit{I}$_2$ and ($ii$) the magnitude of the normal force at local maximum \textit{I}$_3$.
In the case of cyclic (dynamic) movement of the plates, the normal force is positive $F_z > 0$, i.e. the IL keeps pulling apart the plates at point \textit{I}$_2$
and the maximum of the normal force in the point \textit{I}$_3$ ($F^{dyn}_z = 1$~pN) is lower than in the equilibrium case ($F^{eql}_z = 3$~pN).
Both observations indicate that the plate's motion is preventing the ionic liquid to fully fill the void space of the gap.
Also, there is a substantial slip during the ejection of IL from the gap, which results in a lower normal force.
Otherwise, if no slip would be present the maximum normal force at velocity $V_z = 1$~m/s should be about two orders of magnitude higher based on the bulk viscosity coefficient calculated in the previous section.

Partial filling of the gap due to the motion of the walls is even better observable in the results for the interval \textit{II}.
While the equilibrium characteristic has a local maximum, cf. Figure~\ref{fig:fd_static}, in the dynamic case there are only two characteristic points (starting and ending point \{\textit{II}$_1$, \textit{II}$_2$\}
and monotonously increasing normal force between them. At point \textit{II}$_1$ at $d_z = 22$~{\AA} in the static case, we notice that at the mid point between the plates there is a broad maximum of the cation density distribution,
see Figure~\ref{fig:conf_interval2}. In the static case we notice that, similar to the transition from one to two anionic layers within the interval \textit{I}, there is a transition from two to three anionic layers within
the interval \textit{II}, which happens in proximity of the point $d_z = 24$~{\AA}. At point \textit{II}$_2$ we notice two sharp anionic layers in the proximity of the plates and the third anionic layer which is broader,
less sharp and positioned in the middle of the inter--plate gap, cf. Figure~\ref{fig:conf_interval2}.
In the dynamic case the number of layers remains the same in the interval \textit{II}, they just get separated during the extension; and a creation of additional ionic layers by the ions flowing from the lateral reservoirs
into the gap does not take place, cf. Figure~\ref{fig:conf_interval2}.

We can conclude that in a confined system with strong interaction between the walls and the IL, the major driving force that pulls IL into the gap between the plates is the interaction
with the wall atoms rather than the inter--IL interactions.
In order to visualize what happens in the vicinity of the plates, we are presenting snapshots of $xy$ cross--section configurations in the intervals \textit{I} and \textit{II},
check the Figures~\ref{fig:confxyview_interval1} and~\ref{fig:confxyview_interval2}, respectively.
Even on a cursory look, one sees that the phase behaviour of the confined IL is complex: in Figure~\ref{fig:confxyview_interval1}, we observe a salt--like ordering taking place
in all representative points $\left\{I_1, I_2, I_3\right\}$ of the equilibrium configurations. In the dynamic case the IL exhibits some level of ordering for a small gap ($I_1$) and it is amorphous in the other two points.
On the other hand, in Figure~\ref{fig:confxyview_interval2} there was no movement of the IL in and out of the gap and the IL formed a two--dimensional square crystal \{\textit{II}$_1$, \textit{II}$_2$\} on both surfaces
during the dynamic case. In the equilibrium configurations, there are probably enough ions in the gap that allow the IL to obtain its liquid--like character.

At this point, we would like to quantify how could the processes described above contribute to the energy losses. If two macroscopically smooth surfaces come into contact, initially they only touch at a few of these asperity points.
A motion of two bodies in contact lubricated by an ionic liquid would involve the generation of new contacts and the separation of existing ones. Ionic liquids are characterized by strong Columbic interactions between the particles.
By calculating the area covered within the average cycle of the $F_z\left(d_z\right)$ curves in Figure~\ref{fig:fd_dynamic}, we calculate the amount of work invested per average dynamic cycle, i.e., the hysteretic energy losses.
There is a big difference in the amount of invested work in the two intervals: $3.5236$~$\text{pN} \cdot \text{{\AA}}$ for the interval \textit{I} compared to $0.2844$~$\text{pN} \cdot \text{{\AA}}$ for the interval \textit{II},
where the vertical displacement of the Top plate in the two intervals is roughly the same $\Delta d_{z} \approx 5$~{\AA}). This is consistent with a strongly decreasing trend of the maximal normal force
which can be sustained by the system as the number of ionic layers confined between the plates increases, i.e. for the two ionic layers the maximal normal force $F^{\rm I}_{z,max} = 3$~pN,
while for the three ionic layers it is $F^{\rm II}_{z,max} = 0.25$~pN, corresponding to the two maxima of the equilibrium force--distance characteristic in Figure~\ref{fig:fd_static}.

\subsection{Shear Behaviour of Confined Ionic Liquid}
In order to study the behaviour of our confined IL under shearing, we apply a relative motion between the plates along the $x$ direction. The Bottom plate is kept fixed and a constant velocity $V_x$ is imposed on the Top plate.
We are interested in establishing how does the lateral (frictional) force $F_x$ depend on the confinement gap $d_z = \left\{12, 14, 16, 18, 22, 25\right\}$~{\AA}.

\begin{figure}[htb]
\includegraphics[width = 8 cm]{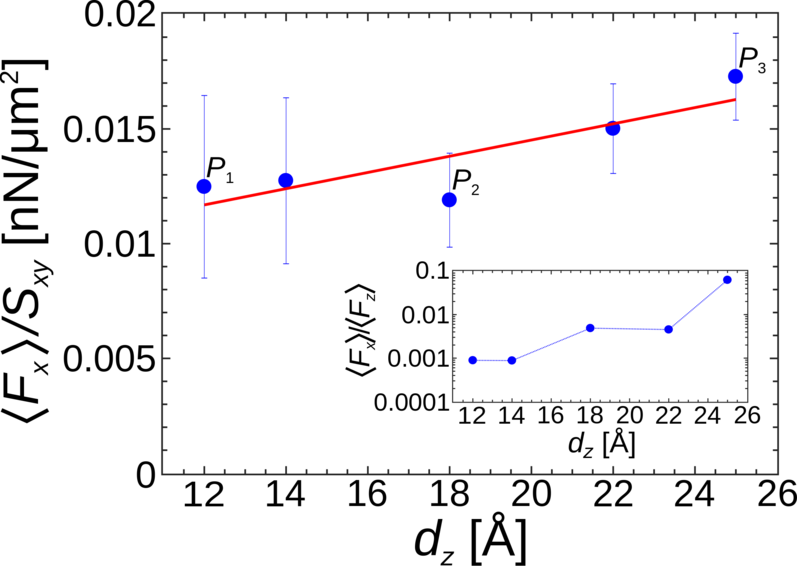}
\caption{Dependence of the frictional force divided by the contact area of the Top plate with IL lubricant $\langle F_x \rangle / S_{\rm xy}$
on the interplate distance $d_z$. The three representative points $\left\{P_1, P_2, P_3\right\}$ are marked.
Points obtained in simulations are shown as circle markers, accompanied with errors along the $y$ axis.
Linear fit through those points is shown as a solid line. In the inset dependence of specific friction $\langle F_x \rangle / \langle F_z \rangle$
on the interplate distance $d_z$ is shown, with $y$ axis in log scale. Simulation points are shown as circle markers, while the dashed line serves as a visual guide.
}
\label{fig:friction_gap_ylog}
\end{figure}

\begin{figure}[htb]
\includegraphics[width = 8 cm]{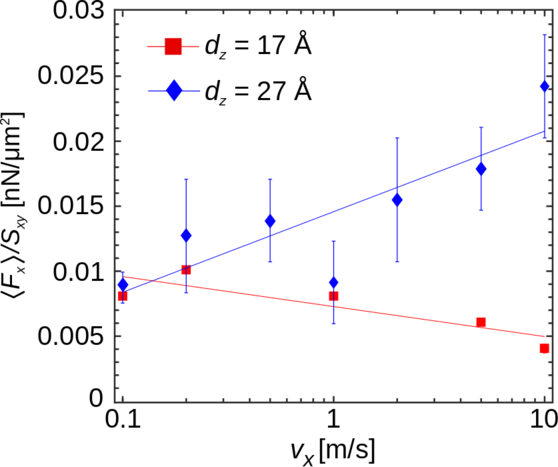}
\caption{Dependence of the frictional force divided by the contact area of the Top plate with IL lubricant $\langle F_x \rangle / S_{\rm xy}$ on the Top plate's lateral velocity $V_x = 0.1-10$~m/s.
The error bars represent the standard deviation of the average values obtained from the simulation data. The lines showing the friction trends are obtained by linear regression.
}
\label{fig:friction_velocity}
\end{figure}

\begin{figure*}[htb]
\includegraphics[width = 16 cm]{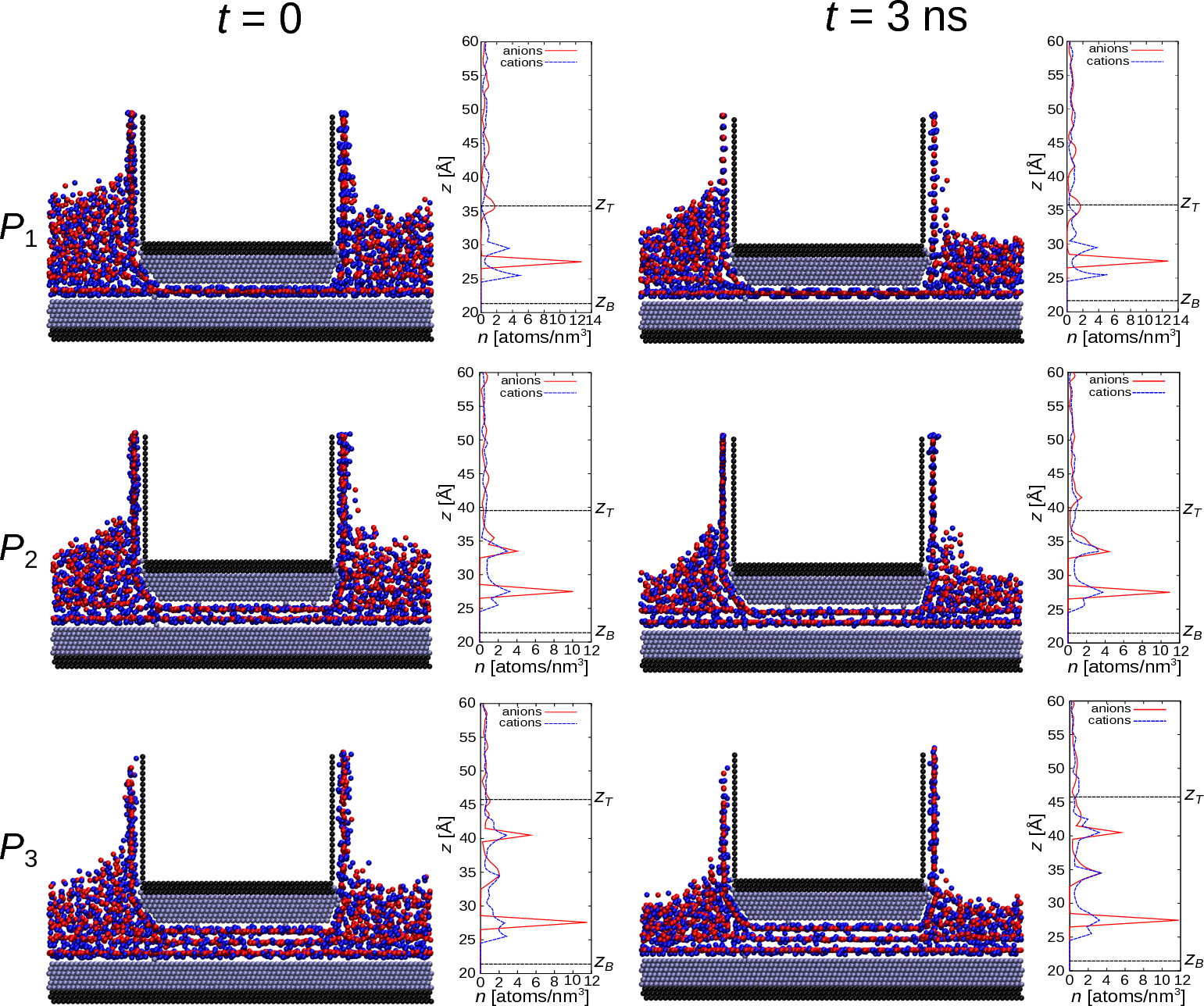}
\caption{Configuration snapshots ($yz$ cross--section) accompanied with ionic density distribution along the $z$ direction in three representative points $\left\{P_1, P_2, P_3\right\}$.
Left panels correspond to the start of friction simulations $t = 0$, while right panels correspond to the end of friction simulations $t = 3$~ns.
Top plate's lateral velocity is set to $V_x = 2$~m/s, total simulation time is $t_{\rm tot} = 3$~ns,
hence all friction simulations have run until the Top plate had covered a distance of $d_x = V_x \cdot t_{\rm tot} = 60$~{\AA} along the $x$ direction.
}
\label{fig:confionicdistribution_FGD}
\end{figure*}

\begin{figure}[htb]
\includegraphics[width = 8 cm]{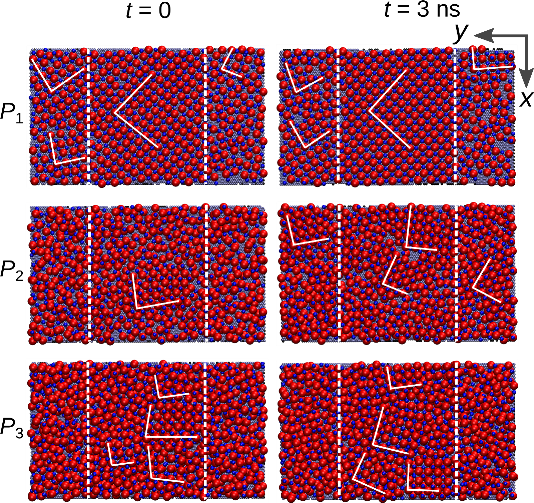}
\caption{Configuration snapshots ($xy$ cross--section) in three representative points $\left\{P_1, P_2, P_3\right\}$.
Left panels correspond to the start of friction simulations $t = 0$, while right panels correspond to the end of friction simulations $t = 3$~ns.
We have highlighted the confined region with dashed lines (Top plate's width along the $y$ axis is a half of the total system's width) and also we have sketched crystallization patterns with solid lines.
Top plate's lateral velocity is set to $V_x = 2$~m/s, total simulation time is $t_{\rm tot} = 3$~ns,
hence all friction simulations have run until the Top plate had covered a distance of $d_x = V_x \cdot t_{\rm tot} = 60$~{\AA} along the $x$ direction.
}
\label{fig:conf_xyview_FGD}
\end{figure}

In Figure~\ref{fig:friction_gap_ylog} we are showing the dependence of the time averaged frictional force divided by the contact area of the Top plate and the IL lubricant, i.e. $\langle F_x \rangle / S_{\rm xy}$
on the interplate distance $d_z$. We observe a linear increase of the frictional force per contact area with the increase of the interplate distance, with a slope of $4$~$\text{nN} / \mu\text{m}^{\text{3}}$.
In the inset of Figure~\ref{fig:friction_gap_ylog}, we are showing the dependence of specific friction defined as the ratio of the time averaged frictional and normal force $\langle F_x \rangle / \langle F_z \rangle$
on the interplate distance $d_z$. By comparing the Figure~\ref{fig:friction_gap_ylog} with the results for the bulk liquid in Figure~\ref{fig:tau_shear_etaGK} we observe that there is no correlation with
the lubricant viscosity (i.e., otherwise frictional force would be three orders of magnitude higher). This leads us to the assumption that our pressurized systems, whether they form a crystalline lattice or not,
do not lie in a typical hydrodynamic regime and operate under full slip conditions in which the ionic liquid moves together with one of the walls. As there is no solid--solid contact between the two surfaces, but lubrication
through very thin, highly viscous films which are solid--like, mixed or dry lubrication are the two potential regimes that can describe the observed conditions.
A parametric study on different shearing velocities $V_x = 0.1-10$~m/s at two wall separations $d_z = 17, 27$~{\AA} provides additional information
for the characterization of the tribological regime of our system. In  Figure~\ref{fig:friction_velocity} one can observe a logarithmic (weak) dependence of the frictional force per contact area
on lateral velocity of the Top plate's movement which is consistent with the observations of previous studies of IL lubrication, cf. Refs.~\cite{mendonca2013ILmetal,CanovaC4CP00005F}.

From Figure~\ref{fig:friction_gap_ylog} we have selected three representative points with $d_z = \left\{12, 18, 25\right\}$~{\AA} labeled as $\left\{P_1, P_2, P_3\right\}$ respectively.
We provide an overview of the $yz$ configuration cross--sections together with ionic density distributions along the $z$ axis (cf. Figure~\ref{fig:confionicdistribution_FGD}) at the simulation onset $t = 0$ and after $t = 3$~ns.
In the panels of Figure~\ref{fig:conf_xyview_FGD} we have highlighted the confined region with dashed lines (the Top plate's width along the $y$ axis is half of the total system's width) and we have also sketched
crystallization patterns with solid lines. In Figures~\ref{fig:confionicdistribution_FGD} and~\ref{fig:conf_xyview_FGD} we show initial configurations at the input of friction simulations,
together with the final configurations obtained after the friction simulations.
We observe that any initial crystallization is not lost due to the lateral motion of the Top plate, but only slightly modified due to the motion, which suggests that the lateral movement does not alter the ordering.
This is a significant finding since the longitudinal movement does alter the local order (it destroys the crystal structure for small gaps and induces it in larger ones).

\section{Conclusions}
In the current work we have used a molecular dynamics simulation setup in order to study the response of a model ionic liquid to imposed mechanical deformation.
The properties of bulk and confined ionic liquid have been investigated under both static and dynamic conditions.
First, we have shown that the Green--Kubo viscosity coefficient fits the shearing simulation results of our bulk salt model ionic liquid, indicating its liquid state.
Our simulation results have shown the significant impact of the confinement and interaction with the walls on the ionic liquid response to mechanical deformation.
The force--distance hysteresis surface under cyclic loading is smaller than one would expect considering only the viscosity value of the liquid.
The simulations have also shown the transition from a liquid to a highly dense and ordered, potentially solidified state of the IL taking place under variable normal load and under shear.
The wall slip has a profound influence on all the forces which arise as a response to the mechanical deformation.
We also observe that the interaction of the IL with the walls represents a principal driving force for all processes observed in the dynamic regime for a range of studied velocities.
If sufficient time is allowed for the system to reach equilibrium, inter--ionic interactions pull more ionic liquid inside the confinement gap.

Ionic liquids feature strong long--ranged Coulombic forces and their models require significant computational effort.
Coarse grained models, such as the salt model implemented in the current study, are useful for bridging the gap between the molecular processes that control the lubrication phenomena and the macroscopic performance
in engineering applications. The implementation of simplified models that describe fundamental physicochemical phenomena at a reduced computational cost can provide deep insights which shed light onto the mechanisms
and processes that can render ILs as potentially interesting lubricant candidates.

\section{Acknowledgements}
The work of M.D. and I.S. was supported in part by the Serbian Ministry of Education, Science and Technological Development under Project No. OI171017 and by COST Action MP1305.
Numerical simulations were run on the PARADOX supercomputing facility at the Scientific Computing Laboratory of the Institute of Physics Belgrade.

\section*{Author Contributions}
K.G. and I.S. designed the study. M.D. and I.S. performed the simulations. They also wrote the paper with inputs from K.G..

\section*{References}

\bibliography{main}

\end{document}